\documentclass{pasj00}

\draft

\begin{document}
\SetRunningHead{S. Kato}{Tidal Instability and Superhump by Wave-Wave Resonant Model}
\Received{2012/00/00}%{yyyy/mm/dd}
\Accepted{2013/12/27}%{yyyy/mm/dd}

\title{Tidal Instability and Superhump in Dwarf Novae by a Wave-Wave Resonant Model} 
%\title{Resonantly Excitaed Disk Oscillations in Two-Armed-Deformed Disks 
%and High-Frequency QPOs in Microquasaras}

%%% begin:list of authors
\author{Shoji \textsc{Kato}}
    %\thanks{Example: Present Address is xxxxxxxxxx}}
\affil{2-2-2 Shikanodai-Nishi, Ikoma-shi, Nara, 630-0114}
\email{kato@gmail.com, kato@kusastro.kyoto-u.ac.jp}
%\and
%\author{C-Firstname {\sc C-Familyname}}
%\affil{C-Address of Institute}\email{ccccc@xxx.xxx.xx.xx}
%%% end:list of authors

%%% Please use the following style in case that sorting by 
%%% affilation is impossible. 
%
% \author{%
%   D-Firstname \textsc{D-Familyname}\altaffilmark{1}
%   E-Firstname \textsc{E-Familyname}\altaffilmark{1,2}
% and
%   F-Firstname \textsc{F-Familyname}\altaffilmark{2}}
% \altaffiltext{1}{Address of Institute}
% \email{ddddd@xxx.xxx.xx.xx}
% \email{eeeee@xxx.xxx.xx.xx}
% \altaffiltext{2}{Address of Institute}

%% `\KeyWords{}' always has to be placed before `\maketitle'.
 %Do NOT move this preamble from here!

\maketitle

\begin{abstract} 
On a disk deformed to a non-axisymmetric form, a set of oscillations
can be excited by their resonant interaction through the disk deformation (Kato et al. 2011).
This resonant instability process has been proposed to suggest a possible cause of 
the high-frequency quasi-periodic oscillations (HF QPOs) observed in black-hole low-mass X-ray binaries.
In the present paper, we examine whether the above-mentioned wave-wave resonant process
can describe the tidal instability and superhump in dwarf novae.
The results show that the process seems to well describe the observations.
If this process is really the cause of the tidal instability and superhump, a two-armed
oscillation with high frequency roughly on the magnitude of three times the orbital frequency is  
present on disks, although its expected amplitude may be small.    
\end{abstract}

\section{Introduction}

The origins of superoutbursts and superhumps observed in dwarf novae are now well understood by
the so-called tidal instability and precession of the eccentric disk deformation 
(Osaki 1985, Whitehurst 1988a,b; Hirose \& Osaki 1990; Lubow 1991a,b, 1992, 1994:
for review see Osaki 1996).
The eccentric disk deformation and its precession were  
found by numerical simulations by Whitehurst (1988a,b).
The origin of the disk deformation was pointed out to be due to a parametric resonance between 
disk rotation and binary revolution from  
a test-particle approximation (Hirose \& Osaki 1990) and later in the framework of hydrodynamics
(Lubow 1991a).
The precession of the deformation is found to be due to an one-armed global oscillation
(Osaki 1985; Hirose \& Osaki 1993).

In a different field of astrophysics, i.e., in the field of black-hole low-mass X-ray binaries (BH LMXBs), 
we know that high-frequency quasi-periodic oscillations (HF QPOs) whose frequencies are in the range of
100 to 450 Hz are observed in some sources (e.g., for review, see van der Klis 2004 and
Remillard \& McClintock 2006).
The origin of these HF QPOs has been studied extensively, since clarification of the
origin will give a powerful tool to know the innermost structure of relativistic accretion disks as well
as the spin of the central black hole sources.
Inspite of many efforts there is still no consensus on the origin of HF QPOs.
However, one of possible models of HF QPOs is excitation of a set of
oscillations by their resonant coupling through disk deformation
(Kato et al. 2011). 
In this model a deformation of unperturbed disks from an axisymmetric state is essential, and 
it is a kind of catalizer for excitation of oscillations.

The above-mentioned process of excitation of disk oscillations is rather general.
Hence, a natural question is whether the tidal instability and superhump
in dwarf novae can be interpreted as a result of excitation of disk oscillations 
on tidally deformed disks by the above-mentioned wave-wave resonant process.
The purpose of this paper is to demonstrate this possibility and 
to obtain some hints on refining the models of the HF QPOs.

\section{Outline of Wave-Wave Resonant Excitation Process through Disk Deformation}
 
Let us assume that oscillations in disks can be decomposed into normal modes.
The time and angular dependences of the displacement vector, $\mbox{\boldmath $\xi$}(\mbox{\boldmath $r$}, t)$,
associated with the oscillations are factorized as
$\mbox{\boldmath $\xi$}(\mbox{\boldmath $r$}, t)=\hat{\mbox{\boldmath $\xi$}}(r,z)
{\rm exp}[i(\omega t-m\varphi)]$.
Here, $\mbox{\boldmath $r$}$ is the cylindrical coordinates ($r, \varphi, z$), whose center is
at the disk center and the $z$-axis is the axis of the rotating axis of the disk.
We consider two oscillations.
The set of frequency and azimuthal wavenumber, i.e., ($\omega$, $m$), of each oscillation is
denoted ($\omega_1$, $m_1$) and ($\omega_2$, $m_2$).
Furthermore, we assume that the disk is deformed to non-axisymmetric state with azimuthal 
wavenumber $m_{\rm D}$ and the pattern rotates with frequency $\omega_{\rm D}$,
i.e., the set of ($\omega$, $m$) of the disk deformation is ($\omega_{\rm D}$, $m_{\rm D}$).
In the case of tidal deformation, $\omega_{\rm D}$ and $m_{\rm D}$ are related by
\begin{equation}
        \omega_{\rm D}=m_{\rm D}\Omega^*_{\rm orb},
\label{0}
\end{equation}
where $\Omega^*_{\rm orb}$ is the orbital frequency of the secondary star around the primary star,
observed from the primary star.
To avoid unnecessary complication, $m_{\rm D}$ is taken to be a possitive integer, i.e., 
$m_{\rm D}=1,2,3...$.

The above two oscillations can non-linearly interact through the disk deformation, if the 
following resonant conditions are satisfied, i.e., 
\begin{equation}
    \omega_2=\omega_1\pm\omega_{\rm D}\quad{\rm and}\quad 
    m_2=m_1\pm m_{\rm D},
\label{2.1}
\end{equation}
where the sign of $+$ or $-$ is possible.
Kato et al. (2011) showed that the above two oscillations grow simultaneously if both oscillations
satisfying conditions (\ref{2.1}) overlap in their propagation regions, and 
have opposite signs of $E/\omega$, i.e., $(E_1/\omega_1)(E_2/\omega_2)< 0$.
Here, $E$ is the wave energy defined by, e.g.,
\begin{eqnarray}
   E_1=\frac{1}{2}\omega_1\biggr[\omega_1\langle\rho_0\mbox{\boldmath $\xi$}_1^*
            \mbox{\boldmath $\xi$}_1\rangle
       -i\langle\rho_0\mbox{\boldmath $\xi$}_1^*(\mbox{\boldmath $u$}_0\cdot\nabla)
            \mbox{\boldmath $\xi$}_1\rangle\biggr]    \nonumber   \\
   \sim\frac{\omega_1}{2}\biggr\langle(\omega_1-m_1\Omega)(\xi_{1,r}^*\xi_{1.r}
        +\xi_{1,z}^*\xi_{1,z})\biggr\rangle,
\label{wave-energy}
\end{eqnarray}
where $\mbox{\boldmath $u$}_0(\mbox{\boldmath $r$})$ is the velocity on the unperturbed disk,
i.e., $\mbox{\boldmath $u$}_0(\mbox{\boldmath $r$})=(0,r\Omega(r),0)$ and the asterisk shows
the complex conjugate.
The condition, $(E_1/\omega_1)(E_2/\omega_2)< 0$, is roughly equal to
$(\omega_1-m_1\Omega)(\omega_2-m_2\Omega)<0$.\footnote
{In Kato et al. (2011), instead of the condition, $(E_1/\omega_1)(E_2/\omega_2)< 0$, opposite signs
of wave energies of two oscillations, i.e., $E_1E_2<0$, is sometimes emphasized as the resonant instability condition,
since they are interested in
oscillations with positive frequencies, i.e., $\omega_1>0$ and $\omega_2>0$.
In some cases of disk deformations, like the case of the present paper, one of $\omega_1$ and $\omega_2$
becomes negative.
Hence, $(\omega_1-m_1\Omega)(\omega_2-m_2\Omega)<0$ is more general than $E_1E_2<0$.
}
This resonant amplification process is shown schematically in figure 1.

This wave amplification comes from an important general characteristic of coupling terms between two oscillations
through disk deformation.
The characteristic is the commutative relations given by equation (3) of Kato et al. (2011)
or equation (83) of Kato (2008).
Explicit expressions for the coupling terms are
given by equations (5), (6) and (82) of Kato (2008).

%---------------------- Figure 1 -----------------------------------
\begin{figure}
\begin{center}
    \FigureFile(120mm,120mm){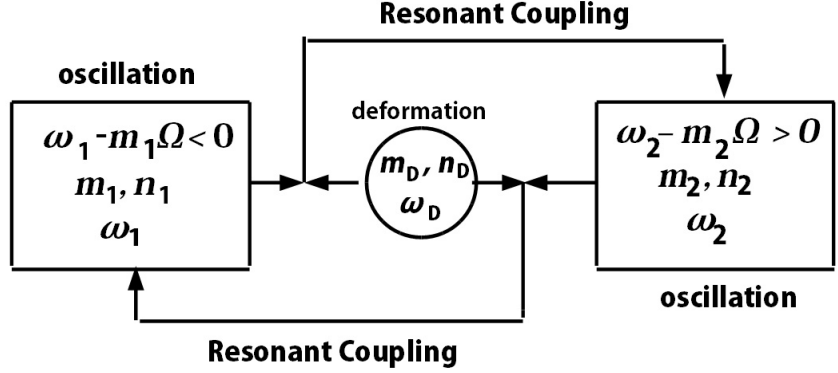}
    %%% \FigureFile(width,height){filename}
\end{center}
\caption{Schematical diagram showing resonant amplification of two oscillations with
opposite signs of $\omega-m\Omega$, i.e., $(\omega_1-m_1\Omega)(\omega_2-m_2\Omega)< 0$
through disk deformation.
The additional necessary conditions for resonance amplification are  
$\omega_2=\omega_1\pm \omega_{\rm D}$, $m_2= m_1\pm m_{\rm D}$, and $n_2=n_1\pm n_{\rm D}$. 
}
\end{figure}
%-------------------------------------------------------------------

It is noted here that for the resonance to be realized an additional condition is necessary.
It is  a relation among node mumbers, say $n$'s, of oscillations in the vertical direction.
This is related to the fact that in the normal mode oscillations, oscillations with different
node numbers are orthogonal with a certain weighting function.\footnote{
For vertically isothermal disks see, for example, Okazaki et al. (1987) and for vertically
polytropic disks see Silbergleit et al. (2001).
}
In the present paper, however, we are only interested in the fundamental mode of oscillations 
in the vertical direction, i.e., $n_1=n_2=0$.
Furthermore, the tidal wave is assumed to have no node in the vertical direction, i.e., $n_{\rm D}=0$.
Since all $n$'s are zero, such an additional condition is automatically satisfied and  
no additional condition is necessary here.

\section{Application to Tidally Deformed Disks}

\subsection{Tidally Deformed Disks}
 
In the case of dwarf novae, the disk of the primary (central) star is deformed by tidal force
of a secondary star. 
The time-averaged part of the tidal potential, ${\bar{\Phi}}$, is given by
\begin{equation}
         {\bar{\Phi}}(r)=-\frac{GM_{\rm s}}{4D^3}r^2,
\label{3.1}
\end{equation}
where $r$ is the radial distance from the central star, $D$ is the binary separation, and $M_{\rm s}$
is the mass of the secondary star.
If the pressure force is neglected, the angular velocity of rotation of the disk gas, $\Omega(r)$,
is given by
\begin{equation}
    \Omega^2r=\frac{GM}{r^2}\biggr(1-\frac{1}{2}q\frac{r^3}{D^3}\biggr),
\label{3.2}
\end{equation}
where $M$ is the mass of the central star and $q=M_{\rm s}/M$.
Since we are interested in the case where $q$ is smaller than unity and in the radial region of
$r/D<1$, we have approximately
\begin{equation}
       \Omega=\Omega_{\rm K}\biggr(1-\frac{1}{4}q\frac{r^3}{D^3}\biggr),
\label{3.3}
\end{equation}
where $\Omega_{\rm K}$ is the Keplerian angular velocity of rotation given by $\Omega_{\rm K}(r)=
(GM/r^3)^{1/2}$. 
In this disk, the epicyclic frequency defined by 
\begin{equation}
     \kappa^2=2\Omega\biggr(2\Omega+r\frac{d\Omega}{dr}\biggr)
\label{3.4}
\end{equation}
is approximately given by
\begin{equation}
     \kappa=\Omega_{\rm K}\biggr(1-q\frac{r^3}{D^3}\biggr).
\label{3.5}
\end{equation}

In addition to the above-mentioned axisymmetric deformation from the Keplerian one, the disk of the 
central star is deformed by non-axisymmetric tidal waves.
The set of ($\omega_{\rm D}$, $m_{\rm D}$) of the deformation is given by 
($m_{\rm D}\Omega^*_{\rm orb}$, $m_{\rm D}$).

\subsection{One-Armed Low-Frequency Oscillation with $\omega-m\Omega<0$}

Since the disk rotation is slightly  deviated from the Keplerian one by the tidal force [see
eq. (\ref{3.3})], an one-armed ($m=1$) p-mode oscillation on the disk is global and has a low frequency
(Osaki 1985, see also Kato 1983).
The dispersion relation for local oscillations [see, e.g., Kato 2001 and Kato et al. 2008]
shows that the radial propagation region of such low frequency oscillation 
[the frequency and azimuthal wavenumber of the oscillation are are denoted $\omega_1$ and $m_1(=1)$, respectively]
is specified by
\begin{equation}
       \omega_1<\Omega-\kappa.
\label{3.6}
\end{equation}
Since $\Omega-\kappa$ is given by [see equations (\ref{3.3}) and (\ref{3.5})]
\begin{equation} 
    \Omega-\kappa=\frac{3}{4}q\Omega_{\rm K} \biggr(\frac{r}{D}\biggr)^3,
\label{3.7} 
\end{equation}
and increases outwards, 
the propagation region of the oscillation with frequency $\omega_1$ is bounded inside.
The inner edge, $r_{\rm c}$, of the propagation region where $\omega_1=\Omega-\kappa$ is  given by 
\begin{equation}
        \omega_1=(\Omega-\kappa)_{\rm c}= \frac{3}{4}q\Omega_{\rm K}(r_{\rm c}) \biggr(\frac{r_{\rm c}}{D}\biggr)^3.
\label{3.8} 
\end{equation}
The outer edge of the propagation region is the outer edge of the disk, $r_{\rm t}$,
which will be specified as the radius where the disk is truncated by wave-wave resonant instability.

The low-frequency one-armed oscillation with frequency $\omega_1$ is thus trapped in the radial
region of $r_{\rm c}<r<r_{\rm t}$.
The region is schematically shown in figure 2.
It is noted that in this propagation region, $\omega_1-m_1\Omega$ is negative, i.e., $\omega_1-m_1\Omega<0$.

\subsection{High-Frequency Oscillation with $\omega-m\Omega>0$ and Efficiency of Coupling}

As the counterpart of the $\omega_1$-oscillation described above, we consider here
an oscillation with $\omega-m\Omega>0$ which satisfies  the resonance conditions (\ref{2.1}).
The oscillation is taken to be a p-mode.
Its frequency and azimuthal wavenumber are denoted, respectively,  $\omega_2$ and $m_2$,
and are determined later.
A p-mode oscillation with given $\omega_2$ and $m_2$ has two propagation regions of 
$\omega_2>m_2 \Omega+\kappa$ and $\omega_2<m_2 \Omega-\kappa$ (e.g., Kato 2001, Kato et al. 2008).
In the former propagation region we have $\omega_2-m_2\Omega>0$, while in the latter we have 
$\omega_2-m_2\Omega<0$.
Hence a p-mode oscillation in the former region is our concern here.

As will be found later, the resonant instability occurs for oscillations with $m_2=-2$.
Thus, $m_2\Omega+\kappa<0$, and as is shown schmatically in figure 2, the $\omega_2$-oscillation 
has $\omega_2<0$ and is trapped between the inner edge, $r_{\rm s}$, of
the disk and the radius $r_{\rm L}$ (the Lindblad resonance) specified by
\begin{equation}
       \omega_2=m_2\Omega_{\rm L}+\kappa_{\rm L},
\label{3.9}
\end{equation}
where the subscript L denotes the value at $r=r_{\rm L}$.
Outside of $r_{\rm L}$ the oscillations are spatially damped.
For trapping to occur, a relation among $r_{\rm s}$, $r_{\rm L}$, and $\omega_2$ is 
necessary as a trapping condition.
In the present problem, however, we need not to pay particular attention
on such a condition by the following reason.
Near to $r_{\rm s}$, the $\omega_2$-oscillation has very short wavelength in the radial direction,
since the difference between $\omega_2$ and $(m_2\Omega+\kappa)_{\rm L}$ becomes large there 
and the difference must be compensated by the radial wavelength becoming short
(consider the dispersion relation of p-mode oscillations.)
In other words, the oscillation with $\omega_2$ is one of high ovetones and near to frequency $\omega_2$
there are many eigen-frequencies densely.
This means that in practice any $\omega_2$ can become a eigen-frequency and
no relation between $r_{\rm s}$, $r_{\rm L}$, and $\omega_2$ is unnecessary here.

For the resonant coupling between two oscillations with ($\omega_1$, $m_1$) and ($\omega_2$, $m_2$)
through disk deformation to occur, the resonant conditions (\ref{2.1}) are necessary as mentioned
before.
In addition, for the coupling to be efficient, the position of $r_{\rm L}$ must be in 
the propagation region of the $\omega_1$-oscillation,
since the coupling efficiency is determined by volume integrations of some linear products of 
displacement vectors $\mbox{\boldmath $\xi$}_1$, $\mbox{\boldmath $\xi$}_2$, and 
$\mbox{\boldmath $\xi$}_{\rm D}$, they being displacement vectors associated with the $\omega_1$- and 
$\omega_2$-oscillations and with the $\omega_{\rm D}$ disk deformation (Kato et al. 2011).
For the coupling terms to become large, i) the region where $\mbox{\boldmath $\xi$}_1$ has a
large amplitude and that where $\mbox{\boldmath $\xi$}_2$ has a large amplitude must overlap,  
and ii) both $\mbox{\boldmath $\xi$}_1$ and $\mbox{\boldmath $\xi$}_2$ in the overlapped region do not
vary in the radial direction with short wavelength.
The latter requirement comes from the fact that if $\mbox{\boldmath $\xi$}_2$, for example, 
changes in the radial direction with short wavelength a volume integration of products among
$\mbox{\boldmath $\xi$}_1$, $\mbox{\boldmath $\xi$}_2$, and $\mbox{\boldmath $\xi$}_{\rm D}$
becomes small by cancellation.
The above consideration concerning the coupling efficiency suggests that the
resonant couplng occurs most strongly in the case of $r_{\rm L}\sim r_{\rm t}$ by the following reasons.
The amplitude of the low-frequency one-armd oscillation is large around $r=r_{\rm t}$
(see section 5 and also figure 3 by Hirose \& Osaki 1993).
Hence, if $r_{\rm L}>r_{\rm t}$, the $\omega_2$-oscillation has short wavelength around $r\sim r_{\rm t}$
and the coupling term resulting from the volume integration becomes small.
On the other hand, if $r_{\rm L}<r_{\rm t}$, the amplitude of the $\omega_2$-oscillation is
small around $r\sim r_{\rm t}$, since the region around $r_{\rm t}$ is the evanescent region
of the $\omega_2$-oscillation.
Based on these considerations, we adopt $r_{\rm L}=r_{\rm t}$ and 
take\footnote{
A rigorous way to know the relation among $r_{\rm L}$, $r_{\rm t}$ and other quantities 
is to calculate the coupling terms and the resulting growth rate of 
oscillations for various sets of $r_{\rm L}$, $r_{\rm t}$ and $r_{\rm c}$ and to combine this procedure
with that determining $r_{\rm t}$ and $r_{\rm c}$ in the following sections.
This is, however, beyond the purpose of this paper.
}   
\begin{equation}
      \omega_2=m_2\Omega_{\rm t}+\kappa_{\rm t},
\label{3.10}
\end{equation}
where the subscript t denotes the value at $r=r_{\rm t}$.

\section{Equation Describing Tidal Truncation Radius $r_{\rm t}$}

As resonant conditions which lead to realistic cases, we adopt 
\begin{equation}
      \omega_2=\omega_1-\omega_{\rm D} 
      \quad {\rm and} \quad  m_2=m_1-m_{\rm D}.
\label{4.1} 
\end{equation}
Then, substituting $\omega_1=(\Omega-\kappa)_{\rm c}$, $m_1=1$, $\omega_{\rm D}=m_{\rm D}\Omega^*_{\rm orb}$,
and equation (\ref{3.10}) into the
first relation of equations (\ref{4.1}), we have as the condition of wave-wave resonant instability
\begin{equation}
    (m_{\rm D}-2)\Omega_{\rm t}=m_{\rm D}\Omega^*_{\rm orb}-(\Omega-\kappa)_{\rm t}
                           -(\Omega-\kappa)_{\rm c},
\label{4.2}
\end{equation}
where the subscript c denotes the value at $r=r_{\rm c}$, and $(\Omega-\kappa)_{\rm t}$ and
$(\Omega-\kappa)_{\rm c}$ are obtained from equation (\ref{3.7}):
\begin{eqnarray}
     (\Omega-\kappa)_{\rm c}=\frac{3}{4}q\Omega_{\rm K}(r_{\rm c}) \biggr(\frac{r_{\rm c}}{D}\biggr)^3, \nonumber \\
     (\Omega-\kappa)_{\rm t}=\frac{3}{4}q\Omega_{\rm K}(r_{\rm t}) \biggr(\frac{r_{\rm t}}{D}\biggr)^3.
\label{4.2'}
\end{eqnarray}
The growth rate of the wave-wave resonant instability depends on magnitudes of coupling terms and disk 
deformation (for a general expression for growth rate, see Kato et al. 2011).

If the difference between $\Omega$ and $\kappa$ is neglected, equation (\ref{4.2}) gives
\begin{equation}
     \Omega_{\rm t}=\frac{m_{\rm D}}{m_{\rm D}-2}\Omega^*_{\rm orb}.
\label{4.3}
\end{equation}
This is the results known as the parametric resonance, i.e., the 3 : 1 resonance in the case of 
$m_{\rm D}=3$, and the 2 : 1 resonance in the case of $m_{\rm D}=4$.
It is noted that in the above cases of $m_{\rm D}=3$ and $m_{\rm D}=4$,
both of $\omega_2$ and $m_2$ are negative.

If we have a relation  between $r_{\rm t}$ and $r_{\rm c}$, equation (\ref{4.2}) gives $r_{\rm t}$ and $r_{\rm c}$
(and thus $\omega_1$ and $\omega_2$) as functions of $q$.
The relation between $r_{\rm t}$ and $r_{\rm c}$ is obtained by considering that the $\omega_1$-oscillation
is trapped between $r_{\rm c}$ and $r_{\rm t}$.
That is, the relation is obtained by imposing the trapping condition, which is a relation among $r_{\rm c}$, 
$r_{\rm t}$ and acoustic speed, $c_{\rm s}$.
Thus, parameters to solve equation (\ref{4.2}) are $q$ and $c_{\rm s}$.

%---------------------- Figure 2 -----------------------------------
\begin{figure}
\begin{center}
    \FigureFile(80mm,80mm){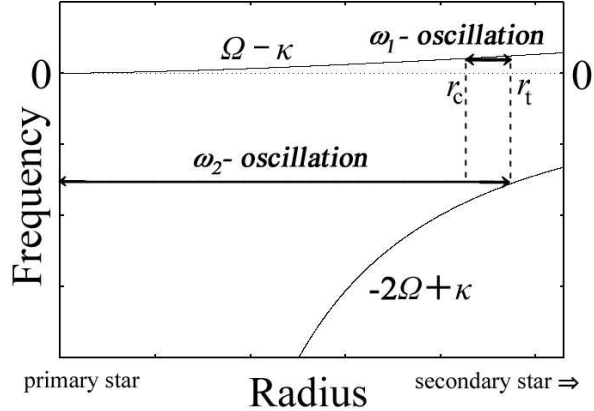}
    %%% \FigureFile(width,height){filename}
\end{center}
\caption{Schematic diagram showing frequencies and propagation regions of the $\omega_1$- and 
$\omega_2$-oscillations. 
The scales of coordinates are arbitrary, and are not linear.
The $\omega_1$-oscillation is trapped between $r_{\rm c}$ and $r_{\rm t}$.
The inside of $r_{\rm c}$ is the evanescent region.
The $\omega_2$-oscillation has a negative frequency and propagates between the inner edge of the disk and $r_{\rm L}$.
(Here, $r_{\rm L}$ and $r_{\rm t}$ are taken to be equal.)
The propagation region is rather wide, but this does not mean that the $\omega_2$-oscillation 
is observed with large amplitude, since the 
oscillation has very short radial wavelength except near to $r_{\rm t}$. 
}
\end{figure}
%-------------------------------------------------------------------
 
\section{Trapping of One-Armed Low-Frequency Global Oscillation}
  
The one-armed low-frequency p-mode oscillation with frequency $\omega_1$ is trapped in the region of 
$r_{\rm c} < r < r_{\rm t}$.
The purpose here is to derive a relation between $r_{\rm c}$ and $r_{\rm t}$ by solving the eigenvalue problem,
introducing acoustic speed in the disk.
This problem has already been examined by Hirose \& Osaki (1993) by using vertically integrated hydrodynamical
equations.
In different contexts, similar problems have been examined, i.e.,
by Okazaki (1991) for V/R variations of Be stars, and by Silbergleit et al. (1990) and others for c-mode
oscillations in relativistic disks.
Here, we consider the simplest situations that the radial wavelength
of perturbations (on the order of $r_{\rm t}-r_{\rm c}$) is so short that the characteristic radial 
scale of variations of unperturbed quantities
in disks can be neglected except when we consider the radial variation of $\omega-(\Omega-\kappa)$.
In this case the wave equation describing the trapped oscillation is (see Appendix)
\begin{equation}
     \biggr(\frac{d^2}{d r^2}+Q \biggr)u_r=0,
\label{5.1}
\end{equation}
where
\begin{equation}
     Q(r)=\frac{(\omega-\Omega)^2-\kappa^2}{c_{\rm s}^2}.
\label{5.2}
\end{equation}

Equation (\ref{5.1}) is now solved by the WKBJ method.
In the region of $r_{\rm c}< r < r_{\rm t}$, the approximate solution can be written as
\begin{equation}
         u_r=Q^{-1/4}{\rm exp}\biggr[\pm i\int Q^{1/2}dr\biggr].
\label{5.3}
\end{equation}
The radius $r_{\rm c}$ is the turning point of $Q$, where $Q=0$.
Near $r_{\rm c}$, the solution of equation (\ref{5.1}) thus can be expressed in terms of the Bessel functions
of the order of $\pm 1/3$ (Morse \& Feshbach 1953).
The asymptotic dependence of the solution for $r\gg r_{\rm c}$ can be arranged so  that it can be expressed
in the form of equation (\ref{5.3}).
Furthermore, by taking only the solution whose amplitude spatially damps in the region of $r< r_{\rm c}$
(the evanescent region of oscillations), we have
(Morse \& Feshbach 1953)
\begin{equation}
   u_r\sim Q^{-1/4}\biggr[{\rm cos}\biggr(w-\frac{5}{12}\pi\biggr)+{\rm cos}\biggr(w-\frac{1}{12}\pi\biggr)\biggr],
\label{5.4}
\end{equation}
where
\begin{equation}
    w(r)=\int_{r_{\rm c}}^r Q^{1/2}dr.
\label{5.5}
\end{equation}

Next, we impose a boundary condition at $r=r_{\rm t}$.
Since in this paper $r_{\rm t}$ is taken to be the outer edge of the disks, the vanishing of the Lagrangian variation of pressure,
i.e., $\delta p=0$, will be relevant.
This is approximately equal to $h_1=0$ (see Appendix for definition of $h_1$), 
and thus to $du_r/dr=0$ at $r=r_{\rm t}$.
From the differentiation of the terms in the brackets of equation (\ref{5.4}) with repect to $r$, we have
(see also Silbergleit et al. 2001)
\begin{equation}
   w_{\rm t}\equiv \int_{r_{\rm c}}^{r_{\rm t}} Q^{1/2}dr =\biggr(n+\frac{1}{4}\biggr)\pi
\label{5.6}
\end{equation}
as the condition determining the wave trapping,\footnote{
If we adopt $u_r=0$ at $r=r_{\rm t}$ as the boundary condition, we have
$$
   w_{\rm t}\equiv \int_{r_{\rm c}}^{r_{\rm t}} Q^{1/2}dr =\biggr(n+\frac{3}{4}\biggr)\pi.  \nonumber
$$
}
where $n=0,1,2,3...$.
Since we are considering the fundamental mode in the radial direction, we adopt hereafter $n=0$,
and $w_{\rm t}=\pi/4$.

We now perform the integration of $Q^{1/2}$ by using expression (\ref{5.2}) for $Q$.
Considering that $(\omega-\Omega)^2-\kappa^2\sim -2\Omega[\omega-(\Omega-\kappa)]$, and further that 
$\omega=\omega_1=(\Omega-\kappa)_{\rm c}$ and $\Omega-\kappa$ is given by equation (\ref{3.7}), we have
\begin{equation}
    Q=\frac{3}{2}\biggr(\frac{GM}{Dc_{\rm s}^2}\biggr)\frac{q}{D^2}\biggr[1-\biggr(\frac{r_{\rm c}/D}
         {r/D}\biggr)^{3/2}\biggr].
\label{5.7}
\end{equation}
Here, we consider that $(r_{\rm t}-r_{\rm c})/r_{\rm c}\ll 1$
(the WKBJ method is still valid since $GM/Dc_{\rm s}^2$ is a large quantity).
Then, we can approximately perform the integration of equation (\ref{5.5}) to have
\begin{equation}
    w_{\rm t}=\biggr(\frac{GM}{Dc_{\rm s}^2}q\biggr)^{1/2}
        \frac{(r_{\rm t}/D)^{3/2}-(r_{\rm c}/D)^{3/2}}
        {(r_{\rm c}/D)^{1/2}},
\label{5.8}
\end{equation}
which is changed to
\begin{equation}
   \frac{r_{\rm t}}{D}=\frac{r_{\rm c}}{D}+\frac{2}{3}
    \frac{w_{\rm t}}{q^{1/2}} \frac{c_{\rm s}}{(GM/D)^{1/2}}.
\label{5.9}
\end{equation}
This is a relation between $r_{\rm t}/D$ and $r_{\rm c}/D$ with dimentionless parameters 
$c_{\rm s}/(GM/D)^{1/2}$ and $q$.

\section{Numerical Calculations}

Substitution of equations (\ref{4.2'}) and (\ref{5.9}) into equation (\ref{4.2}) gives 
the disk truncation radius $r_{\rm t}$ as a function of $q$ and $c_{\rm s}/(GM/D)^{1/2}$.
The $q$ - $r_{\rm t}$ relation for $c_{\rm s}/(GM/D)^{1/2}=0.02$ is shown in figure 3.
The $q$ - $r_{\rm t}$ relation depends little on $c_{\rm s}/(GM/D)^{1/2}$.
The $q$ - $r_{\rm c}$ relation, on the other hand, depends on $c_{\rm s}/(GM/D)^{1/2}$,
and the relation in three cases of $c_{\rm s}/(GM/D)^{1/2}=0.01$, 0.02,
and 0.04 are also shown in figure 3.
The region between $r_{\rm t}$ and $r_{\rm c}$ is the trapped one of the $\omega_1$-oscillation.
As is expected, the width increases with increase of $c_{\rm s}/(GM/D)^{1/2}$.

%---------------------- Figure 3 -----------------------------------
\begin{figure}
\begin{center}
    \FigureFile(80mm,80mm){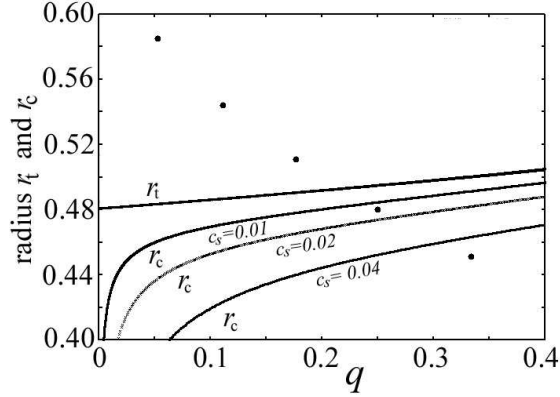}
    %%% \FigureFile(width,height){filename}
\end{center}
\caption{The $q$ - $r_{\rm t}/D$ and $q$ - $r_{\rm c}/D$ relations calculated by our model.
The former relation is drawn for $c_{\rm s}/(GM/D)^{1/2}=0.02$, which is almost free from
the value of $c_{\rm s}/(GM/D)^{1/2}$.
The latter relation is drawn for three cases of $c_{\rm s}/(GM/D)^{1/2}=0.01$, 0.02, and 0.04.
The region between $r_{\rm t}$ and $r_{\rm c}$ is the trapped region of the $\omega_1$-oscillation.
The trapped region becomes wide as $c_{\rm s}$ increases.
For comparison, the maximum radial size of periodic orbit of a test particle in the binary potential 
calculated by Pacz\'{n}ski (1977) is ploted for some values of $q$ by black circles.
The tidal instability occurs for $q< 0.22$.
}
\end{figure}
%-------------------------------------------------------------------

Paczy\'{n}ski (1977) estimated the maximum size of accretion disks in a binary system by
calculating the maximum size of periodic orbit of a test particle.
One of such radii, e.g., $r_{\rm max}$ in his  paper, is also shown
in figure 3, for comparison, for some values of $q$.
The comparison shows that the disk size is limitted by the tidal instability in the case where 
$q< 0.22$.

The observed superhump frequency $\omega_{\rm sh}$ is related to $\Omega_{\rm orb}$ and $\omega_1$
by (Osaki 1985)
\begin{equation}
       \omega_{\rm sh}=\Omega_{\rm orb}-\omega_1,
\label{6.1}
\end{equation}
where $\Omega_{\rm orb}$ is the binary orbital frequency in the inertial frame [i.e., $\Omega_{\rm orb}
= \Omega_{\rm orb}^*(1+q)^{1/2}$].
Then, the dimensionless quantity defined by $\epsilon\equiv P_{\rm sh}/P_{\rm obr}-1$,
where $P_{\rm sh}$ and $P_{\rm orb}$ are the period of superhump and the orbital period
(in the inertial frame), respectively, is  written as
\begin{equation}
     \epsilon=\frac{\omega_1/\Omega_{\rm orb}}{1-\omega_1/\Omega_{\rm orb}}.
\label{6.2}
\end{equation}
Here, in our present model, $\omega_1/\Omega_{\rm orb}$ is given by [see eq. (\ref{3.8})]
\begin{equation}
    \frac{\omega_1}{\Omega_{\rm orb}}\sim\frac{3}{4}\frac{q}{(1+q)^{1/2}}\biggr(\frac{r_{\rm c}}{D}\biggr)^{3/2}.
\label{6.3}
\end{equation}
The $q$ - $\epsilon$ relation calculated by using the above relations (\ref{6.2}) and (\ref{6.3}) 
is drawn in figure 4 for
three cases of $c_{\rm s}/(GM/D)^{1/2}= 0.01$, 0.02 and 0.04.
In order to compare with observations, the curves are superposed on the $q$ - $\epsilon$
plot by T.Kato et al.(2012) for recently observed sources.
The calculated curve in the case of $c_{\rm s}/(GM/D)^{1/2}=0.02$, which is $c_{\rm s}/(D\Omega^*_{\rm orb})$,
seems to be close to observational results.
The dimensionless value of $c_{\rm s}/(GM/D)^{1/2}=0.02$ will be relevant observationally, since
in accretion disks in cataclysmic variables, $D\Omega_{\rm orb}\sim 500{\rm km}\ {\rm s}^{-1}$
and $c_{\rm s}\sim 10{\rm km}\ {\rm s}^{-1}$.

%---------------------- Figure 4 -----------------------------------
\begin{figure}
\begin{center}
    \FigureFile(80mm,80mm){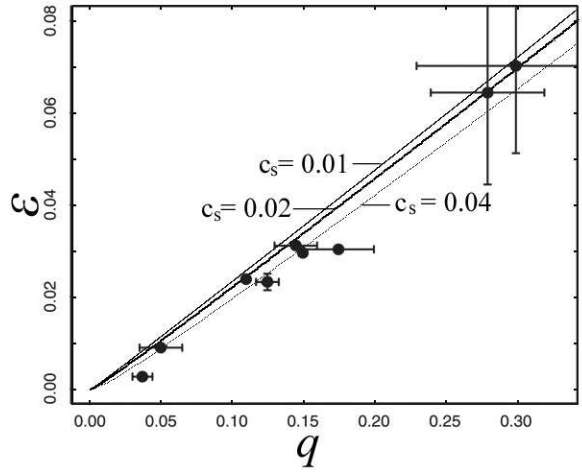}
    %%% \FigureFile(width,height){filename}
\end{center}
\caption{The $q$ - $\epsilon$ relation calculated for three cases of $c_{\rm s}/(GM/D)^{1/2}=$ 
0.01, 0.02, and 0.04.
These curves are superposed on the diagram plotting observed $q$ - $\epsilon$ relation 
(figure 34 of T. Kato et al. 2009).
The names of nine sources in this figure are listed in table 6 of T.Kato et al. (2009). }
\end{figure}
%-------------------------------------------------------------------

\section{Discussions}

The present wave-wave resonant excitation model well describes the tidal instability - superhump
phenomena in dwarf novae.
This is not surprising, since the present excitation model can be regarded just as a combination 
of the tidal instability model by Lubow (1991a) and the precession of low frequency eccentric
deformation by Osaki (1985) into a unified scheme. 
%As long as the author knows, the theory of tidal instability 
%seems not to describe explicitly why the instability makes the eccentric deformation rotate.
%Osaki (1985) and Hirose \& Osaki (1993) demonstrate the presence of a slowly precessing eccentric oscillation mode,
%but it does not show why the oscillation is excited in the stage of the tidal instability.
%The present wave-wave resonant excitation  model is a bridge between the above two.
In the tidal instability model by Lubow (1991a), the precession of the disk deformation is not essential in understanding
the essence of the instability.
The tidal torque and pressure force, however, make the deformation precess (Lubow 1992, 1994).
Hence, there is a connection between the instability and eccentric precession, and much attention has been taken to 
the connection (Goodchild \& Ogilvie 2006, Ogilvie 2007, Lubow 2010).
In these studies, the temperature dependence of the precession has been examined (Lubow 2010).
Here, we compare our results with those by Lubow (2010).
Figure 5 shows the $\omega_1$ - $(H/r)_{\rm t}$ relation in the case of $q=0.1$, where
$H$ is the disk half-thickness and taken simply to be $c_{\rm s}/\Omega_{\rm K}$ and the subscript t denotes
the value at the disk truncation radius, $r_{\rm t}$.
The results show that the precession rate decreases with increase of $(H/r)_{\rm t}$, which agrees with results by
Lubow (see figure 6 of Lubow 2010).
Quantitatively, however, there are some differences.
These differences partially come from differences of situations considered.
In Lubow's calculations the disk edge is taken at $0.5D$, while in our calculations it is taken at $r_{\rm t}$ 
which is determined by relation (\ref{4.2}) and smaller than $0.5D$, depending on $(H/r)_{\rm t}$.
In Lubow's calculations the rate of precession is taken to become the gravitational precession rate of a free particle on an 
excentric orbit in the limit of no temperature, while in our calculations it is taken to tend to $(\Omega-\kappa)_{\rm t}$.

It is noted that in our wave-wave resonant instability model two small amplitude oscillations are superposed on a steady 
deformed disk, and their resonant instability is considered.
In other words, in a stage where the disk deformation is growing rapidly with time, the model cannot be applied.
One of other important approximations involved in the present model is that $r_{\rm t}$ and $r_{\rm L}$ are assumed to be
equal.
In the limit of no temperature, the $\omega_2$-oscillation is localized sharply at $r_{\rm L}$ (the radius of the 
inner Lindblad resonance), and thus the coupling constants between the $\omega_1$- and $\omega_2$- oscillations
are large only when $r_{\rm t}$ and $r_{\rm L}$ are equal.
In disks with finite temperature, however, the radial region where the coupling constants are large becomes
wide.
As is mentioned in a footnote, to know a proper relation between $r_{\rm t}$ and $r_{\rm L}$ we must calculate the
radial distribution of the coupling coefficients, using the functional forms of the oscillations.
This is a complicated problem.
This situation corresponds to the fact that in disks with finite temperature the resonant region has a finite width
(e.g., Meyer-Vernet \& Sicardy 1987).

%---------------------- Figure 5 -----------------------------------
\begin{figure}
\begin{center}
    \FigureFile(80mm,80mm){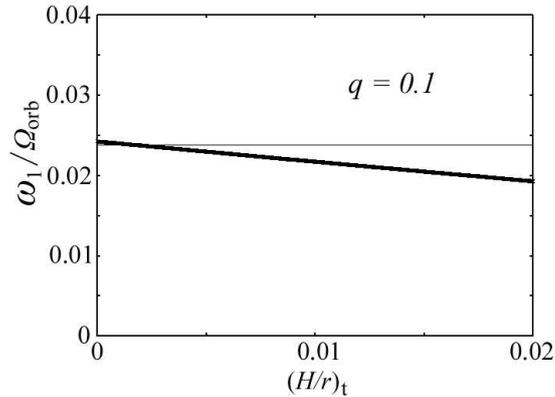}
    %%% \FigureFile(width,height){filename}
\end{center}
\caption{The temperature dependence of $\omega_1$ in the case of $q=0.1$.
The rate of precession, $\omega_1$, is normalized by $\Omega_{\rm orb}$ and the effects of temperature are 
measured by $(H/r)_{\rm t}$, where $H$ is the disk half-thickness and $H/r$ is taken to be $c_{\rm s}/\Omega_{\rm K}$.
In the limit of zero temperature, $\omega_1$ tends to $(\Omega-\kappa)_{\rm t}$ in our model. 
For comparison, $\Omega-\kappa$ at the radius of $\Omega=3\Omega^*_{\rm obs}$ is shown by a thin line.
}
\end{figure}
%-------------------------------------------------------------------
 
In our model the presence of two oscillation modes with opposite signs of $\omega-m\Omega$
are essential for instability, and 
one of the oscillations excited (i.e., one-armed low-frequency oscillation with
$\omega-m\Omega<0$) brings about a precession of the deformation.
The other oscillation with $\omega-m\Omega>0$ is not directly related to the superhump, 
but important for resonant instability.
This latter mode corresponds to the intermediate disk deformation mode in Lubow's resonant feedback
process (Lubow 1991a).
In our present model, $\vert m_2\vert = 2$ like Lubow's case, 
and its frequency $\vert\omega_2\vert$ changes with the change of $\omega_1$, since they are related by a resonance condition 
as $\vert\omega_2\vert=\omega_{\rm D}-\omega_1$.
The frequency $\vert\omega_2\vert$ of the oscillation is shown in figure 6 as a function of $q$
in the case of $c_{\rm s}/(GM/D)^{1/2}=0.02$.
The curve depends little on the value of $c_{\rm s}/(GM/D)^{1/2}$, 
but changes in the range of $3.0\sim 2.4\Omega_{\rm orb}$ for change of $q$ in the range of $0<q< 0.4$.
 
Observational detection of the $\omega_2$-oscillation and its q-dependence is an interesting subject 
to judge whether the
present wave-wave resonant instability model represents real situations of tidal instability - superhump phenomena.
Osaki (2003) pointed out that complex superhump light curves for the 2001 outburst of WZ Sagittae can be interpreted 
as excitation of many oscillation modes.
Especially, he emphasizes the appearance of two armed oscillation with frequency $3\Omega_{\rm orb}-\omega_1$, and considers it 
as a support of Lubow' mode-coupling model.
Except for this case, however, there seems to be no observational evidence which suggests the 
coexistence (with the $\omega_1$-oscillation) of the $\omega_2$-oscillation 
in the superoutburst stage (private communication by T. Kato).
Osaki suggests that this is due to the fact that many sources with superhumps are observed pole-on.
In addition there will be following situations for difficulty of observations of the $\omega_2$-oscillation.
As mentioned before, the $\omega_2$-oscillation has short wavelength except in the region near to $r_{\rm t}$,
although its propagation region is wide.
Hence, a large luminosity variation will not be expected, since various phases of the oscillation are superposed
on observed quantities.
Furthermore, $r_{\rm L}=r_{\rm t}$ adopted in this paper is an approximation.
The outer part of disks will be fluctuating with time in general,
and this will introduce time fluctuations in the relation between $r_{\rm L}$ and $r_{\rm t}$.
If so, the phase and frequency of the $\omega_2$-oscillation will be subject to it, more than those of
the $\omega_1$-oscillation.
This will smear observational properties of the $\omega_2$-oscillation.

%---------------------- Figure 6 -----------------------------------
\begin{figure}
\begin{center}
    \FigureFile(80mm,80mm){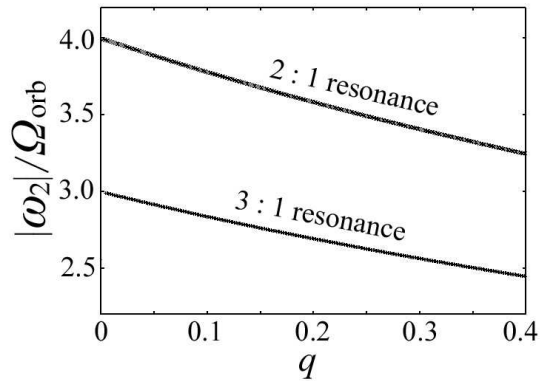}
    %%% \FigureFile(width,height){filename}
\end{center}
\caption{The $q$ - $\vert\omega_2\vert$ relation calculated for $c_{\rm s}/(GM/D)^{1/2}=0.02$.
The corresponding curves in the cases of $c_{\rm s}/(GM/D)^{1/2}=0.01$, and 0.04 are almost the same
as that in the case of $c_{\rm s}/(GM/D)^{1/2}=0.02$.
For reference, the frequency $\vert\omega_2\vert$ in the case of the 2 : 1 resonance is also drawn as a function
of $q$ for $c_{\rm s}/(GM/D)^{1/2}=0.02$.
}
\end{figure}
%-------------------------------------------------------------------

So far, we have considered the so-called 3 : 1 resonance which occurs for $m_{\rm D}=3$.
In this case the $\omega_2$-oscillation is a two-armed one (i.e., $\vert m_2\vert=2$) 
with frequency close to $3\Omega_{\rm orb}$ as mentioned above.
For tidal deformation of $m_{\rm D}=4$, on the other hand, the instability occurs near 2 : 1 resonance
[see equation ({\ref{4.3})].
In this case, $r_{\rm t}$ is larger than that in the case of the 3 : 1 resonance.
The $\omega_2$-oscillation in this case is three-armed 
(i.e., $\vert m_2\vert=3$) and its 
frequency is close to $4\Omega_{\rm orb}$ [see equation (\ref{4.1})].
As mentioned before, observational detection of the $\omega_2$-oscillation is interesting 
to evaluate validity of the present wave-wave resonant model.
Thus, for reference, the frequency, $\vert \omega_2\vert$, in the case of the 2 : 1 resonance is
also drawn in figure 5 as a function of $q$.

We think that the high frequency quasi-periodic oscillations (HF QPOs) in microquasars will result from
a similar process considered here.
Hence, detailed examinations how well the present wave-wave resonant model can describe the 
superoutburst-superhump phenomena in dwarf novae is helpful to consider more the excitation mechanism
of HF QPOs in BH LMXBs.

\bigskip
The author thanks T. Kato for discussions on various observational aspects of superhumps, and 
Y. Osaki for pointing out the presence of high frequency oscillation of $\Omega_{\rm orb}-\omega_1$
in an outburst stage of WZ Sagittae.
The author also thanks the referee for pointing out recent theoretical developments on disk precession
and for many helpful suggestions.  

\bigskip\noindent
{\bf Appendix:  Wave Equation Describing Disk Oscillations}

For simplicity, the disk is assumed to be vertically isothermal.
Then, the vertical hydrostatic balance gives
\begin{equation}
    \rho_0(r,z)=\rho_{00}(r){\rm exp}\biggr(-\frac{z^2}{2H^2}\biggr),
\label{A1}
\end{equation}
where $\rho_0(r,z)$ is the density in the unperturbed disk and $\rho_{00}(r)$ is that on the equatorial
plane.
The scale height in the vertical direction, $H$, is given by
\begin{equation}
     H^2(r)=\frac{c_{\rm s}^2}{\Omega_{\rm K}^2}\biggr(1+\frac{1}{2}q\frac{r^3}{D^3}\biggr)
           \equiv \frac{c_{\rm s}^2}{\Omega_\bot^2},
\label{A2}
\end{equation}
where $c_{\rm s}(r)$ is the isothermal  acoustic speed and $\Omega_\bot$ is the vertical epicyclic 
frequency in the tidally deformed disks.

On the disks we impose small-amplitude isothermal perturbations.
The perturbations are assumed to be proportional to exp$[i(\omega t-m\varphi)]$.
Then, hydrodynamical equations describing the small-amplitude perturbations are
\begin{eqnarray}
   i(\omega-m\Omega)\rho_1+\frac{\partial}{r\partial r}(r\rho_0u_r)-i\frac{m}{r}\rho_0u_\varphi   \nonumber \\
        +\frac{\partial}{\partial z}(\rho_0 u_z)=0,
\label{A3}
\end{eqnarray}
\begin{equation}
    i(\omega-m\Omega)u_r-2\Omega u_\varphi=-\frac{\partial h_1}{\partial r},
\label{A4}
\end{equation}
\begin{equation}
    i(\omega-m\Omega)u_\varphi+\frac{\kappa^2}{2\Omega}u_r=i\frac{m}{r}h_1,
\label{A5}
\end{equation}
\begin{equation}
     i(\omega-m\Omega)u_z=-\frac{\partial h_1}{\partial z},
\label{A6}
\end{equation}
where $h_1$ is defined by
\begin{equation}
       h_1=\frac{p_1}{\rho_0}=c_{\rm s}^2\frac{\rho_1}{\rho_0}.
\label{A7}
\end{equation}
Here, ($u_r$, $u_\varphi$, $u_z$) are
velocity perturbations over unperturbed one ($0$, $r\Omega$, 0), 
and $\rho_1$ and $p_1$ are the density and pressure perturbations over the unperturbed ones,
$\rho_0$ and $p_0$, respectively.

Hereafter, we consider perturbations whose radial wavelength is shorter than the characteristic
radial scales of unperturbed quantities in disks.
Elimination of $u_\varphi$ from equation (\ref{A4}) and (\ref{A5}) then gives
\begin{equation}
    [-(\omega-m\Omega)^2+\kappa^2]u_r=-i(\omega-m\Omega)\frac{\partial h_1}{\partial r}.
\label{A8}
\end{equation}
Furthermore, the continuity equation (\ref{A3}) is reduced to
\begin{equation}
   i(\omega-m\Omega)\frac{h_1}{c_{\rm s}^2}+\frac{\partial u_r}{\partial r}
   +\biggr(\frac{\partial}{\partial z}-\frac{z}{H^2}\biggr)u_z=0.
\label{A9}
\end{equation} 
Under the same approximation, elimination of $u_z$ from this equation by using equation (\ref{A6}) gives
\begin{equation}
    \frac{\partial^2h_1}{\partial z^2}-\frac{z}{H^2}\frac{\partial h_1}{\partial z}
      +\frac{(\omega-m\Omega)^2}{c_{\rm s}^2}h_1
      =i(\omega-m\Omega)\frac{\partial u_r}{\partial r}.
\label{A10}
\end{equation}

Equations (\ref{A8}) and (\ref{A10}) are the set of equations to be solved.
As mentioned in the text, we are interested in one-armed ($m=1$) low-frequency oscillations,
where $\omega\sim \Omega-\kappa$.
Hence, the radial variation of $(\omega-\Omega)^2-\kappa^2$ should be taken into account, but
the radial variations of other quantities are neglected.
Then, operating $\partial^2/\partial z^2-(z/H^2)\partial/\partial z+(\omega-m\Omega)^2/c_{\rm s}^2$ on
equation (\ref{A8}) and eliminating $h_1$ from the resulting equation by using (\ref{A10}),
we have an partial differential equation with respect to $u_r$.
After manipulations we have
\begin{eqnarray}
   \biggr(\frac{\partial^2}{\partial z^2}-\frac{z}{H^2}\frac{\partial}{\partial z}\biggr)u_r
   +\frac{(\omega-\Omega)^2}{c_{\rm s}^2}u_r   \nonumber \\
   +\frac{(\omega-\Omega)^2}{(\omega-\Omega)^2-\kappa^2}\frac{\partial^2}{\partial r^2}u_r=0,
\label{A11}
\end{eqnarray}
where $m=1$ has been adopted.

This partial differential equation can be easily decomposed into two ordinary differential equations.
That is, decomposing $u_r(r,z)$ as $u_r(r,z)=g(z)f(r)$, we have
\begin{equation}
    \biggr(\frac{\partial^2}{\partial z^2}-\frac{z}{H^2}\frac{\partial}{\partial z}+K\biggr)g(z)=0
\label{A12}
\end{equation}
and
\begin{equation}
  \biggr[\frac{(\omega-\Omega)^2}{(\omega-\Omega)^2-\kappa^2}\frac{\partial^2}{\partial r^2}
     +\frac{(\omega-\Omega)^2}{c_{\rm s}^2}-K\biggr]f(r)=0,
\label{A13}
\end{equation}
where $K$ is the separation constant, and can be determined by solving equation (\ref{A12}) with
relevant boundary conditions at $z=\pm \infty$.

Equation (\ref{A12}) is the Hermite equation.
Imposing that $\rho_0^{1/2}(z)g$ (which is proportional to kinetic energy
of perturbations) remains finite at $z=\pm \infty$, we find that $K$ should be zero or a positive integer 
and $g(z)$ is the Hermite polynomials (Okazaki et al. 1987):
\begin{equation}
     K = n, \quad n=0,1,2,...
\label{A15}
\end{equation}
and
\begin{equation}
  g(z)={\cal H}_n(z/H).
\label{A14}
\end{equation}

In the present problem, we are interested in the fundamental p-mode oscillation, i.e., $n=0$.
Hence, equation (\ref{A13}) is reduced to
\begin{equation}
     \biggr[\frac{d^2}{dr^2}+\frac{(\omega-\Omega)^2-\kappa^2}{c_{\rm s}^2}\biggr] f(r) =0.
\label{A16}
\end{equation}
This is the equation which we treat in the text.

\bigskip
\leftskip=20pt
\parindent=-20pt
\par
{\bf References}
\par
Goodchild, S. \& Ogilvie, G. I. 2006, MNRAS, 368, 1123\par
Hirose, M., \& Osaki, Y. 1990, PASJ, 42, 135 \par
Hirose, M., \& Osaki, Y. 1993, PASJ, 45, 595 \par
Kato, S. 1983, 35, 249\par
Kato, S. 2001, PASJ, 53, 1\par 
Kato, S. 2008, PASJ, 60, 111 \par
%Kato, S. 2010, PASJ, 62, 635 \par
%Kato, S. 2011a, PASJ, 63, 125 \par
%Kato, S. 2011b, PASJ, 63, 861 \par
%Kato, S. 2011c, PASJ, ??? \par
%Kato, S. \& Fukue, J. 1980, PASJ, 32, 377\par
%Kato, S., Honma, \& F. Matsumoto, R. 1988, MNRAS, 231, 37 \par
%Kato, S., Fukue, J., \& Mineshige, S. 1998, Black-Hole Accretion Disks 
%  (Kyoto: Kyoto University Press), chap. 17 \par
Kato, S., Fukue, J., \& Mineshige, S. 2008, Black-Hole Accretion Disks --- Towards a New paradigm --- 
  (Kyoto: Kyoto University Press), chap. 11 \par
Kato, S., Okazaki, A.T.,\& Oktariani, F. 2011, 63, 363 \par
Kato, T. et al. 2009, PASJ, 61, S395\par
Lubow, S. H. 1991a, ApJ, 381, 259 \par
Lubow, S. H. 1991b, ApJ, 381, 268 \par 
Lubow, S. H. 1992, ApJ, 401, 317\par
Lubow, S. H. 1994, in {\it Theory of Accretion Disks -- 2}, ed., W.J. Duschl, J. Frank, F. Meyer, 
    E. Meyer-Hofmeister, \&  W.M., Tscharnuter (Kluwer Academic Publishers, Dordrecht), p.109 \par
Lubow, S. H. 2010, MNRAS, 406, 2777\par
Meyer-Vernet, N. \& Sicardy, B. 1987, Icarus, 69, 157\par
Morse, P.M., \& Feshbach, H., 1953, Methods of Theoretical Physics, Part II 
    (McGraw-Hill, Inc., New York), p.1092\par 
Ogilvie, G. I. 2007, MNRAS, 374, 131 \par
Okazaki, A.T. 1991, PASJ, 43, 75 \par
Okazaki, A.T., Kato, S., \& Fukue, J. 1987, PASJ, 39, 457\par
%Ortega-Rodriguez, M., Silbergleit, A.S., \& Wagoner, R.V. 2002, ApJ, 567, 1043\par
Osaki, Y. 1985, A\&A, 144, 369 \par 
Osaki, Y. 1996, PASP, 108, 390 \par
Osaki, Y. 2003, PASJ, 55, 841 \par
Paczy\'{n}ski, B. 1977, ApJ, 216, 823\par
%Perez, C.A., Silbergleit, A.S., Wagoner, R.V., \& Lehr, D.E. 1997, ApJ, 476, 589 \par
Remillard, R.A., \& McClintock, J.E. 2006, ARA\&A, 44, 49\par
Silbergleit, A.S., Wagoner, R.V., Ortega-Rodr\'{i}gues, M. 2001, ApJ, 548, 385\par
van der Klis, M. 2004, in Compact Stellar X-ray Sources, eds. W.H.G. Lewin and M. van der Klis 
   (Cambridge Univ. Press, Cambridge), 39\par
Whitehurst, R. 1988a, MNRAS, 232, 35 \par
Whitehurst, R. 1988b, MNRAS, 233, 529 \par

\leftskip=0pt
\parindent=0pt
\bigskip\noindent
Note added on Jan. 3, 2013

In the present paper, we have restricted our attention only to the resonances of
$\omega_2=\omega_1\pm\omega_{\rm D}$ and $m_2=m_1\pm m_{\rm D}$.
If the resonances of $\omega=-\omega_1\pm\omega_{\rm D}$ and $m_2=-m_1\pm m_{\rm D}$ are
considered, the instability condition in the latter cases is found to be 
$(E_1/\omega_1)(E_2/\omega_2)>0$, different from $(E_1/\omega_1)(E_2/\omega_2)<0$ in the former cases of 
$\omega_2=\omega_1\pm\omega_{\rm D}$ and $m_2=m_1\pm m_{\rm D}$.
If the latter resonances are taken into account, we see that in addition to the $\omega_2$-oscillations
considered in this paper (i.e., $\omega_2=\omega_1-\omega_{\rm D}<0$ and $m_2=m_1-m_{\rm D}<0$), the 
$\omega_2$-mode of oscillations with $\omega_2=-\omega_1+\omega_{\rm D}>0$ and
$m_2=-m_1+m_{\rm D}>0$ is also excited.
These two oscillations are, however, the same, since both signs of $\omega$ and $m$ are changed simultaneously.
A detailed re-examination of wave-wave resonant instability by Kato et al. (2011) will be made in a subsequent paper.

\end{document}